\documentclass[oldversion]{aa}  
\usepackage{graphicx}
\usepackage{natbib}

\usepackage{txfonts}
\begin{document}
   \title{Molecular gas in NUclei of GAlaxies (NUGA)}

   \subtitle{VII. NGC\,4569, a large scale bar funnelling gas into the nuclear region\thanks{Based on observations carried out with the IRAM Plateau de Bure Interferometer. IRAM is supported by INSU/CNRS (France), MPG (Germany) and IGN (Spain).}}

   \author{F.\,Boone\inst{1,2}
	\and
	 A.\,J.\,Baker\inst{3}
	\and
	E.\,Schinnerer\inst{4}
	\and
        F.\,Combes\inst{1}
	\and
 	S.\,Garc\'\i a-Burillo\inst{5}
	\and
	R.\,Neri\inst{6}
        \and
	L.\,K.\,Hunt\inst{7}
        \and
	S.\,L\'eon\inst{8}
	\and
	M. Krips\inst{9}
	\and
	L.\,J.\,Tacconi\inst{10}
	\and
	A.\,Eckart\inst{11}
          }

   \offprints{frederic.boone@obspm.fr}

   \institute{LERMA, Observatoire de Paris, 
              61 avenue de l'Observatoire, F-75014 Paris, France
\and
	Max-Planck-Institut f\"ur Radioastronomie, auf dem H\"ugel 69, 53121 Bonn, Germany
\and
	Department of Physics and Astronomy, Rutgers, the State University of New Jersey, 136 Frelinghuysen Road, Piscataway, NJ 08854, USA
\and
	Max-Planck-Institut f\"ur Astronomie,	
	K\"onigstuhl 17,
	D-69117 Heidelberg, Germany
         \and
	Observatorio Astron\'omico Nacional (OAN) - Observatorio de Madrid, C/ Alfonso XII 3, 28014 Madrid, Spain
   	\and	
        Institut de Radio Astronomie Millim\'etrique, 
	300 Rue de la Piscine, 38406 St. Mt. d'H\`eres, France
	\and
        INAF-Istituto di Radioastronomia/Sezione,
	Largo Enrico Fermi 5, 
	50125 Firenze, Italy
	\and
	IRAM Pico Veleta Observatory,
	Avenida Divina Pastora 7, Local 20
	E-18012 Granada, Spain	
	\and
	Harvard-Smithsonian Center for Astrophysics, SMA project, 645 N A`ohoku Pl., Hilo, HI,96720, USA;
	\and
	Max-Planck-Institut f\"ur extraterrestrische Physik, 
	Postfach 1312, 85741 Garching, Germany
	\and	
	Physikalisches Institut, Universit\"at zu K\"oln, 
	Z\"ulpicherstrasse 77, D-50937 K\"oln, Germany
             }

   \date{Received February 7, 2007; accepted May 17, 2007}

 
  \abstract
   {This work is part of the NUGA survey of CO emission in nearby active galaxies. We present observations of NGC\,4569, a member of the Virgo Cluster.
  We analyse the molecular gas distribution and kinematics in the central region and we  investigate a possible link to the strong starburst present at the nucleus. 70$\%$ of the $[1.1\pm0.2]\times10^9$\,M$_{\odot}$ of molecular gas detected in the inner 20$^{\prime\prime}$ is found to be concentrated within the inner 800\,pc and is distributed along the large scale stellar bar seen in near-infrared observations. A hole in the CO distribution coincides with the nucleus where most of the H$\alpha$ emission and blue light are emitted.
The kinematics are modelled in three different ways, ranging from the purely geometrical to the most physical. This approach allows us to constrain progressively the physical properties of the galaxy and eventually to emerge with a reasonable fit to an analytical model of orbits in a barred potential. 
Fitting an axisymmetric model shows that the non-circular motions must be comparable in amplitude to the circular motions (120\,km\,s$^{-1}$). {Fitting a model based on elliptical orbits allows us to identify with confidence the single inner Lindblad resonance (ILR) of the large scale bar}. 
Finally, a model based on analytical solutions for the gas particle orbits in a weakly barred potential constrained by the ILR radius reproduces the observations well. 
The mass inflow rate is then estimated and discussed based on the best fit model solution. The gravitational torques implied by this model are able to efficiently funnel the gas inside the ILR down to 300\,pc, although another mechanism must take over to fuel the nuclear starburst inside 100\,pc.}
   \maketitle
%
%
\section{Introduction}

NUGA, for NUclei of GAlaxies \citep{2003agnc.conf..423G}, is a survey of nearby active galaxies aiming for a deep insight into the mechanisms responsible for feeding the nuclei of active galaxies. High resolution (down to 0.5'') and high sensitivity (rms of $\sim$2\,mJy\,beam\,s$^{-1}$ for 10\,km\,s$^{-1}$ channels) observations of 12 galaxies were obtained with the IRAM  Plateau de Bure Interferometer (PdBI) and the IRAM 30\,m telescope. The CO\,$J$$=$1--0 and 2--1 transitions thus mapped allow us to study the dynamics of the molecular gas in the central kiloparsec of the galaxies with a linear resolution as high as $\sim$10\,pc.

The main results achieved by the NUGA project thus far can be summarized as follows:
\begin{itemize}
\item  A great variety of morphologies and dynamics is observed: 2 and 1-arm spirals \citep{2003A&A...407..485G}, bars, rings \citep{2004A&A...414..857C}, circumnuclear discs, asymmetries, and warps \citep{2005A&A...442..479K}. As a corollary, there is no evidence for features (at scales $>$10\,pc) that would be typical of active galaxies and that could be uniquely linked to the activity of the nucleus.
\item We do not see any 2D-kinematic pattern indicative of systematic inflow motions in the circumnuclear disks of some of the NUGA galaxies examined thus far down to the scales of our spatial resolution (10\,pc at best). In some cases, e.g. NGC\,4826 \citep{2003A&A...407..485G} and NGC\,7217 \citep{2004A&A...414..857C}, the observed perturbations to the disks seem less likely to drive than to inhibit the fueling of the nucleus. 
\item A detailed study of the gravitational torques exerted  by the large scale disk asymmetries  on the gas in four galaxies showed that such asymmetries may not be the dominant mechanism for angular momentum removal in the nuclear regions \citep{2005A&A...441.1011G}.
\end{itemize} 

These first results suggest either that low luminosity AGN (the NUGA sample includes Seyferts and LINERs) do not require efficient fueling to sustain their luminosity \citep[see e.g. ][]{2003ASPC..290..379H}, or that the timescales for fueling and the onset of activity are so  different that both cannot be observed simultaneously \citep[see e.g.][]{2004IAUS..222..383C}. However, more galaxies in the sample need to be analysed before any definitive conclusions about AGN fueling can be drawn.

NGC\,4569 (M90) is a bright SAB(rs)ab galaxy at a distance of {17\,Mpc} \citep{1983A&A...118....4B, 1988ngc..book.....T} in the Virgo Cluster. A large scale bar is seen in NIR images \citep{2002MNRAS.337.1118L} and is almost aligned with the major axis of the galaxy \citep[PA=15\,$\deg$ according to][]{2005ApJ...630..837J}. The galaxy harbours a nucleus of the transition type \citep[type T2 in][]{1997ApJS..112..315H} which exhibits, by far, the most pronounced nuclear starburst activity among the LINERs and transition nuclei with available UV data \citep{1998AJ....116...55M}. As such it became a prime target for studies of the origin of the ionization in transition nuclei. The presence of a supergiant-dominated starburst was established by \citet{1996PASP..108..917K}. \citet{2000PASP..112..753B} demonstrated that a starburst with a large population of very hot Wolf-Rayet stars could produce the observed spectrum of the nucleus and AGN activity is ruled out by X-ray \citep{2000ApJ...539..161T, 2001A&A...380...40T, 2001ApJ...549L..51H} and radio observations \citep{1987A&AS...70..517H, 1992AJ....103.1746N}. \citet{2002AJ....124..737G} showed that for the starburst to be the only source of photoionization (no AGN) extreme conditions are required, with $\sim$5$\times 10^{4}$ O and B stars squeezed into the inner 30\,pc.

On larger scales \citet{2001A&A...380...40T} detected a large soft X-ray emitting region above the disk of NGC\,4569 which implies a large energy input into the halo. \citet{2003IAUJD..10E..38H} reported on a bipolar outflow seen in X-rays and a giant outflow ($\sim$10\,kpc in length) to the western side of the galaxy in H$\alpha$ (see also Bomans et al 2007, submitted). Using the Effelsberg radio telescope at 4.85 GHz and 8.35 GHz, \citet{2006A&A...447..465C} discovered large symmetric lobes of polarized radio emission extending up to 24 kpc from the galactic disk on each side of the nucleus. These lobes may have been powered by the nuclear starburst. 

H\,{\sc i} line observations reveal that the galaxy is H\,{\sc i} deficient compared to galaxies of similar type and it must have lost more than 90\% of its  atomic gas \citep{1988A&AS...72...57W, 1990AJ....100..604C}. This was most likely stripped due to rapid motion of the galaxy through the intracluster medium \citep{2004A&A...419...35V, 2006ApJ...651..811B}.

Interferometric CO observations of NGC\,4569 were presented in \citet{2003ApJS..145..259H},  \citet{2005ApJ...630..837J} and \citet{2005PASJ...57..905N}.
In this article, higher resolution and higher sensitivity CO(1--0) and, for the first time, CO(2--1) line observations are presented, and the analysis is focused on investigating a possible link between the molecular gas kinematics in the inner kiloparsec and the nuclear starburst. The observations are presented in Section\,\ref{sec:obs}. The gas kinematics are modeled based on analytical expressions for the gas orbits. Three different models are compared to the observations starting with a pure geometrical model and proceeding to more realistic models in Section\,\ref{sec:models}. {Comparing the different models to the observations allows us to progressively constrain the kinematic properties of the galaxy and emerge with a reasonable fit.} The mass inflow due to gravitational torques based on the best fit version of the most realistic model is estimated in Section\,\ref{sec:massinflow}. {The goal of this estimate is to quantify the possible contribution of the gravitational torques to the fueling of the nuclear starburst.} The results are discussed in Section\,\ref{sec:discussion} and summarized in Section\,\ref{sec:conclusion}.
\section{CO Observations and global properties of the emission}\label{sec:obs}

\begin{figure*}
  \includegraphics[width=\textwidth]{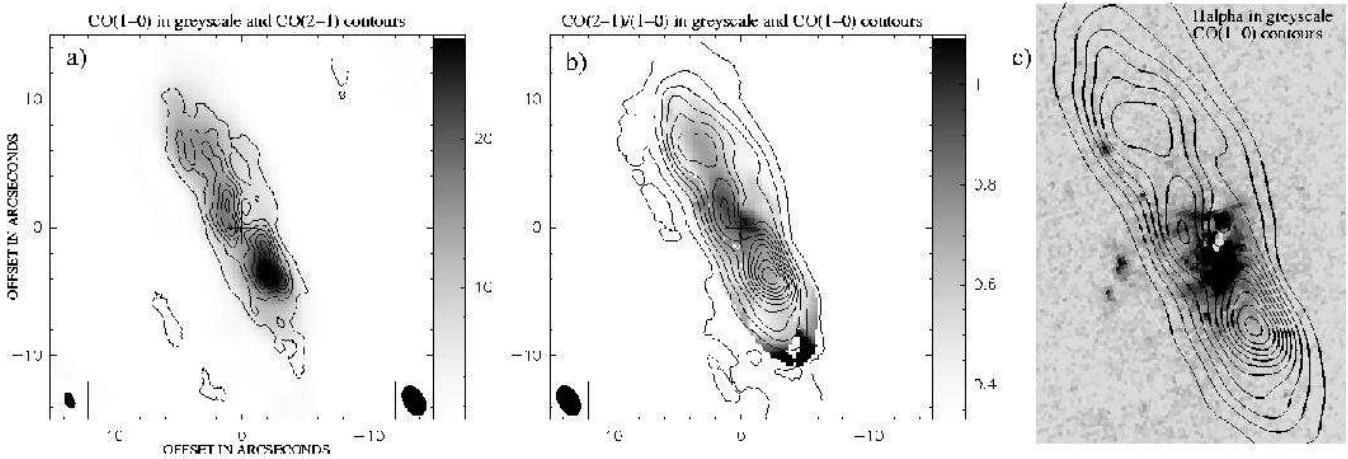}
  \caption{{\bf a)} CO(1--0) (grey scale) and CO(2--1) (contours) integrated maps with a clip at zero and no correction for primary beam attenuation. The color scale is given on the right hand-side in Jy\,beam$^{-1}$\,km\,s$^{-1}$. The contours start at 4\,$\sigma$ and are spaced by 10\,$\sigma$ with $\sigma$=0.38\,Jy\,beam$^{-1}$\,km\,s$^{-1}$. The 115\,GHz and 230\,GHz beams are shown on the bottom right and left respectively. The axes give the angular offset from the phase center (symbolized by the cross) in arcseconds. {We recall that, at 17\,Mpc, 1$''$ corresponds to 82\,pc along the major axis.} {\bf b)} CO(2-1)/CO(1-0) ratio map with the CO(1-0) contours overlaid. The contours start at 4\,$\sigma$ and are spaced by 10\,$\sigma$, where $\sigma$$=$0.27\,Jy\,beam$^{-1}$\,km\,s$^{-1}$.  {\bf c)} HST H$\alpha$ greyscale with CO(1--0) contours.}\label{fig:21_ratio_ha}
\end{figure*}

\begin{figure*}
  \rotatebox{0}{\includegraphics[width=14cm]{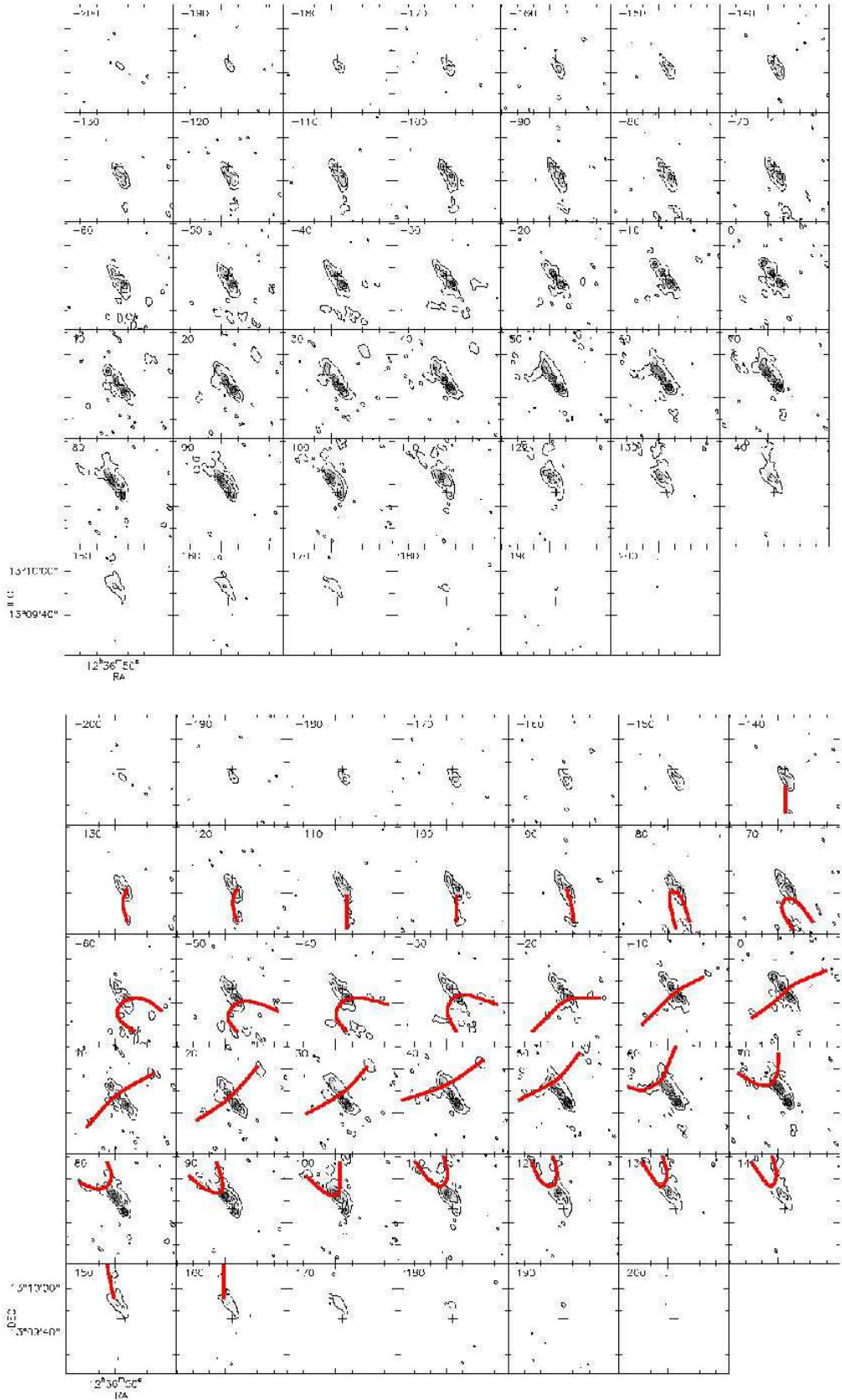}}
  \caption{CO(1--0) channel maps. The velocity relative to the systemic velocity of the galaxy ($v_{\rm hel}$$=$235\,km\,s$^{-1}$) is given at the top left of each map. The phase center symbolized by the cross is at $\alpha_{J2000}$$=$$12^h 36^m 49.8^s$, $\delta_{J2000}$$=$$13^{\circ}09'46.3''$. The beam is 1.8$''$$\times$1.1$''$ and the rms is 2.1\,mJy\,beam$^{-1}$. The contours start at 4\,$\sigma$ and the levels are separated by 20\,$\sigma$. In the bottom panel the channel maps are overlaid with thick lines emphasizing the two ``arms'' seen at low level (see Section\,\ref{sec:veldist}).}\label{fig:co10chann}
\end{figure*}

\begin{figure*}
  \rotatebox{0}{\includegraphics[width=\textwidth]{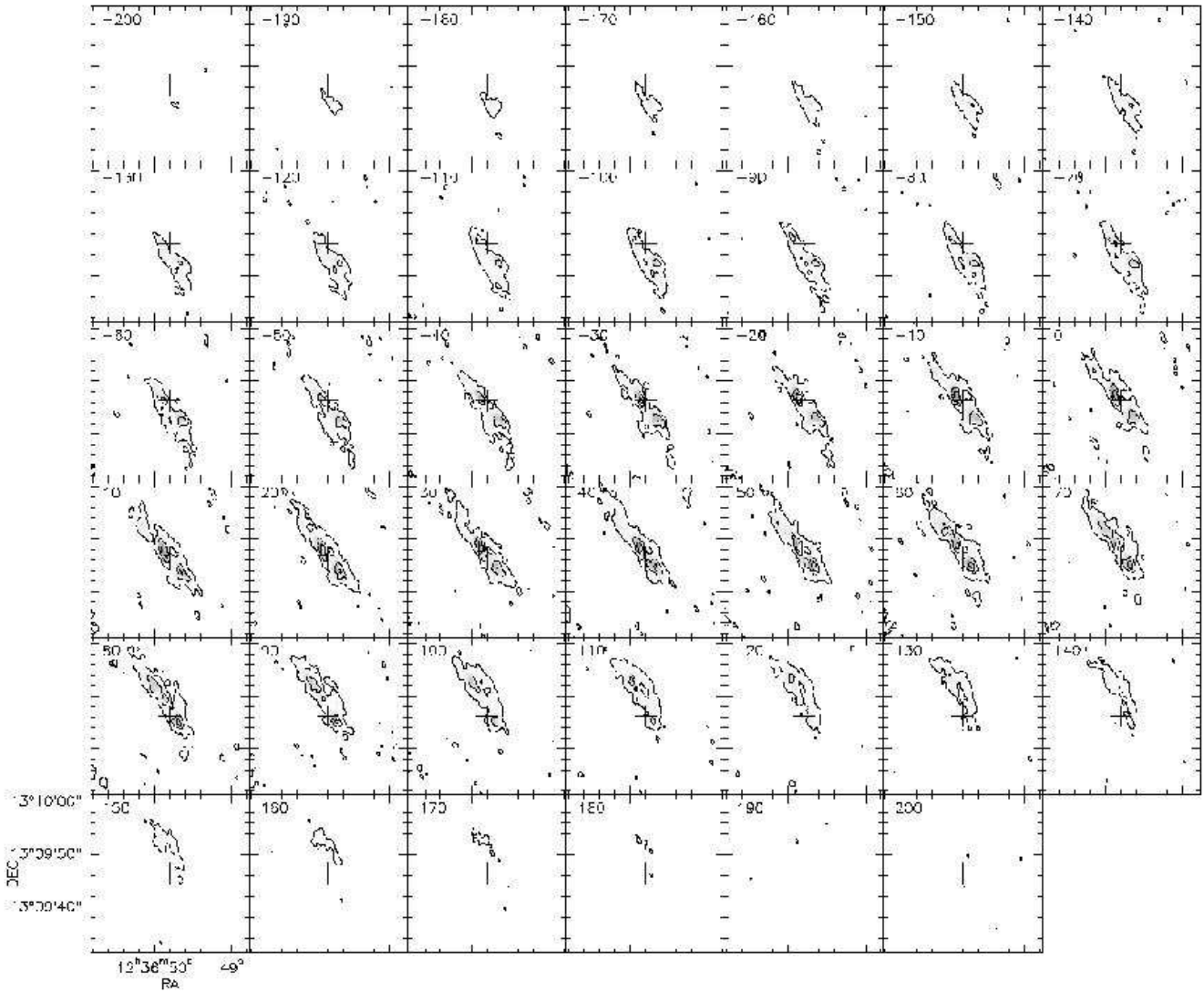}}
  \caption{CO(2--1) channel maps. The beam is 1.1$''$$\times$0.7$''$ and the rms is 3.7\,mJy\,beam$^{-1}$. The contours start at 4\,$\sigma$ and the levels are separated by 20\,$\sigma$.}\label{fig:co21chann}
\end{figure*}
\begin{figure}
\begin{center}
 \rotatebox{0}{\includegraphics[width=\columnwidth]{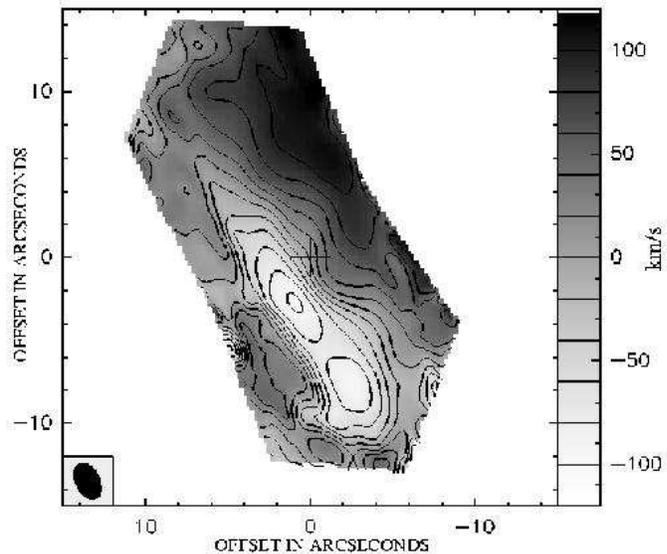}}
\end{center}
  \caption{First moment map of the CO(1--0) emission. Regions with low S/N are masked out. The contour levels are spaced by 10\,km\,s$^{-1}$. The color scale is given on the right hand-side in \,km\,s$^{-1}$.}\label{fig:velomap}
\end{figure}
\subsection{Observations}

The galaxy was observed with the IRAM Plateau de Bure Interferometer (PdBI) in 2003 (B and C configurations) and 2004 (A configuration)  and with the IRAM 30\,m telescope in 2004 in the  CO(1--0) (115\,GHz) and CO(2--1) (230\,GHz) emission lines. The PdBI and 30\,m telescope receiver characteristics, the observing setup and the observing procedures are the same as in \citet{2004A&A...414..857C}. The quasar 3C273 was used for PdBI gain calibration. The maps obtained from the 30\,m observations are used to compute the short spacings and complete the interferometric measurements using the SHORT-SPACE task in the GILDAS software \citep{2000ASPC..217..299G} as in \citet{2004A&A...414..857C}. The RMS in 10\,km\,s$^{-1}$ wide channels is 3.9 and 5.4\,mJy\,beam$^{-1}$ for natural weighting and the beamsizes are 2.5$''$$\times$1.6$''$ and 1.1$''$$\times$0.7$''$ at 115\,GHz and 230\,GHz, respectively.
\subsection{Morphology and mass of molecular gas}
Both CO(1--0) and CO(2--1) emission are  distributed over a region 17$''$$\times$6$''$  in size (Fig.\,1\,a\,\&\,b) roughly aligned with the major axis of the galaxy (PA=24\,deg). The CO(1--0) distribution shows three peaks, one peak close to the center and two symmetrical peaks at $\sim$5'' from the center and aligned with the overall elongation. The south-western peak is the strongest. The central peak does not coincide with the dynamical center: it appears to be the top of a slightly curved ridge encircling  a hole in the CO distribution centered $\sim$1$''$ northwest of the dynamical center. Since our observations have a S/N $\ge$10 times higher than those presented by   \citet{2005PASJ...57..905N}, a weaker continuous distribution is detected in addition to the three main peaks, even when going to the highest resolution in the  CO(2--1) map.

The integrated CO(1--0) flux within a radius of 20$''$ is 450\,Jy\,km\,s$^{-1}$ with an uncertainty due to calibration errors of 15$\%$. This is compatible with the fluxes measured by  \citet{2005PASJ...57..905N} and \citet{2005ApJ...630..837J}. The corresponding H$_2$ mass is  $[1.1\pm0.2]\times10^9$\,M$_{\odot}$ using a conversion factor of $X=2.2\times 10^{20}$cm$^{-2}$[K~km\,s$^{-1}$]$^{-1}$ \citep[][]{1991IAUS..146..235S}. 70$\%$ of this flux is concentrated within a radius of 10$''$ ($\sim$800\,pc). 
\subsection{Velocity distribution}\label{sec:veldist}
The line-of-sight CO velocities span a range of 400\,km\,s$^{-1}$.
Channel maps are shown in Figures\,\ref{fig:co10chann} and \ref{fig:co21chann}.  The most noticeable features are: (1) the distribution is fairly symmetric with respect to systemic velocity and dynamical center; (2) the emission is elongated in most of the channels (from $-$150 to $+$150\,km\,s$^{-1}$) along a direction close to that of the emission in the integrated map; (3) from one channel to the next as velocity increases,  the elongated emission is slightly shifted toward the northwest; (4) the highest velocities ($\pm$200\,km\,s$^{-1}$) are reached at $\sim$5'' from the nucleus, i.e. close to the two outer peaks in the integrated map; and (5) {some low-level emission, mostly seen in the CO(1--0) channel maps  (Fig.\,\ref{fig:co10chann}) outside the elongated core, seems to follow the ``spider'' pattern of rotating disks, i.e. two ``arms'' on each side of the major axis shift from the north at low velocities to the south at high velocities. To guide the eyes of the reader the two arms are emphasized with thick red lines drawn by hand and overlaid to the channel maps in the bottom panel of Fig.\,\ref{fig:co10chann}. Due to the low level of the signal these arms appear discontinuous and the spider pattern is not seen in the velocity field of Fig.\,\ref{fig:velomap} (the first moment mixes this low signal with the noise).}

The fact that in individual channels the emission is elongated in a direction close to that of the major axis {(PA=24\,deg)} implies that significant amounts of gas must lie at ``forbidden'' velocities, i.e. velocities of a given sign appear on both sides of the minor axis, as was noted by   \citet{2005ApJ...630..837J}.
The velocity field is shown in Fig.\,\ref{fig:velomap}; it is clearly distorted compared to a disk with only circular rotation. In the central region a steep velocity gradient is running in a SE to NW direction, i.e. orthogonal to the major axis. {The kinematic line of nodes is therefore orthogonal to the major axis.} This steep gradient results from the continuous shift of the emission in the channel maps (feature (3) described above).

\subsection{Line ratio}

The CO(2--1)/(1--0) line ratio was computed after tapering the 230\,GHz data and restoring it with the same beam as for the 115\,GHz data; the ratio map is shown Fig.\,\ref{fig:21_ratio_ha}\,b. The values are in the range 0.3--1.1 in temperature units. The highest values are reached in the ridge between the central and the southern peaks near the hole in CO. A high line ratio corresponds to high density and/or high temperature gas. A possible interpretation is that the molecular gas in the ridge is being compressed and heated by the nuclear starburst. This interpretation would also be consistent with the presence of a hole in the molecular gas distribution: part of the molecular gas undergoing this pressure might have been dissociated or pushed out of the disk. A fraction of the initial molecular gas in the center must also have been consumed by star formation. The high ratio in the southern peak could be either due to a star forming region or a kinematic shock resulting from the asymmetrically perturbed gravitational potential.

\subsection{Comparison to optical observations}\label{sec:optical}
\begin{figure*}
\begin{centering}
  \rotatebox{0}{\includegraphics[width=\textwidth]{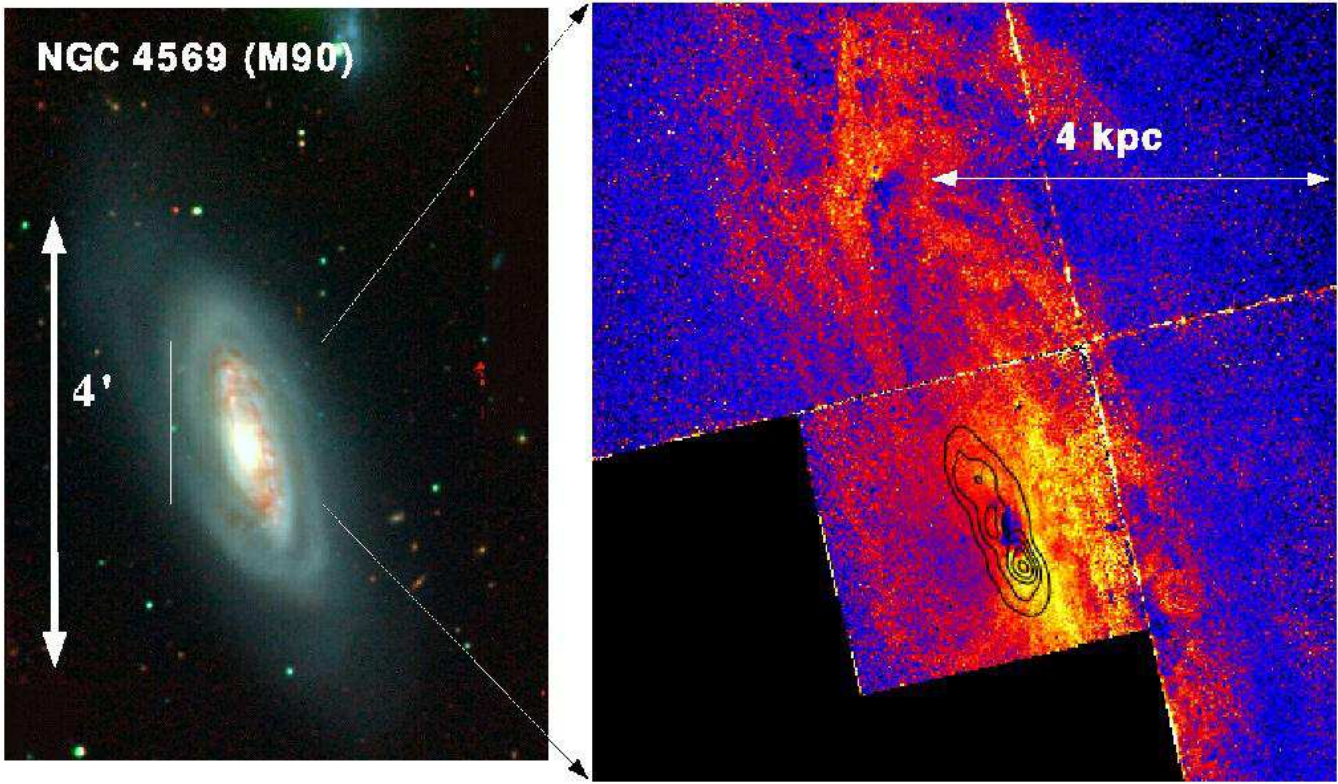}}
  \caption{{\bf Left:} RGB color image synthesized from B,V and H broad band images obtained from Observatoire de Haute Provence (OHP) and retrieved from the Goldmine database \citep{2003A&A...400..451G}. North is up and East is left. {\bf Right:} V-I color index from HST WFPC2-camera observations (F555W and F814W filters, proposal 05375). The color scale is such that high V-I index appears in yellow and low V-I in blue.}\label{fig:dust}
\end{centering}
\end{figure*}
%
%
%
A composite color map from B, V and H broad band images  and the V-I color map from HST observations  are presented in Fig.\,\ref{fig:dust}. Dust lanes are clearly seen along the bar, which in NIR observations extends up to a radius of $\sim$5\,kpc \citep{2002MNRAS.337.1118L}. The bar appears to be aligned with the major axis of the galaxy, like the CO emission, as noted by \citet[][]{2005ApJ...630..837J}. Also striking is the difference in extinction between the eastern and western sides of the galaxy. This confirms that the near side, i.e. the side that appears as the most extinguished because of a low contribution from foreground bulge stars, is the western side. This geometry is consistent with the spiral arms being trailing as in most spiral galaxies. The two outer CO(1--0) peaks coincide with peaks in dust extinction, the southern one being the more prominent both in extinction and CO(1--0) emission. The central CO(1--0) peak does not coincide with any extinction peak, and the hole in the CO(1--0) distribution close to the location of the nucleus has blue optical colors. This agreement suggests that most of the starburst is concentrated within a region of $\sim$100\,pc in radius, supporting the hypothesis that the hole in the  CO(1--0) emission is a consequence of this concentrated starburst having consumed or expelled the molecular gas from the center.  The latter interpretation is further supported by an HST H$\alpha$ image \citep{2000ApJ...532..323P} which shows strong emission south of the hole and several features elongated toward the east (Fig.\,\ref{fig:21_ratio_ha}\c). Some H$\alpha$ emission is expected due to the ionizing stars of the starburst but part of it might also trace the impact of the starburst on the ISM and the elongated features to the east are suggestive of an outflow perpendicular to the plane. The east side being the far side, the counter-outflow (i.e. emerging from the other side of the plane) might be obscured and therefore not visible in  H$\alpha$, although again visible further to the west as reported by \citet{2003IAUJD..10E..38H}.

%
%
%
\section{Kinematic modeling}\label{sec:models}
\begin{figure*}
\begin{center}
  \rotatebox{0}{\includegraphics[width=\textwidth]{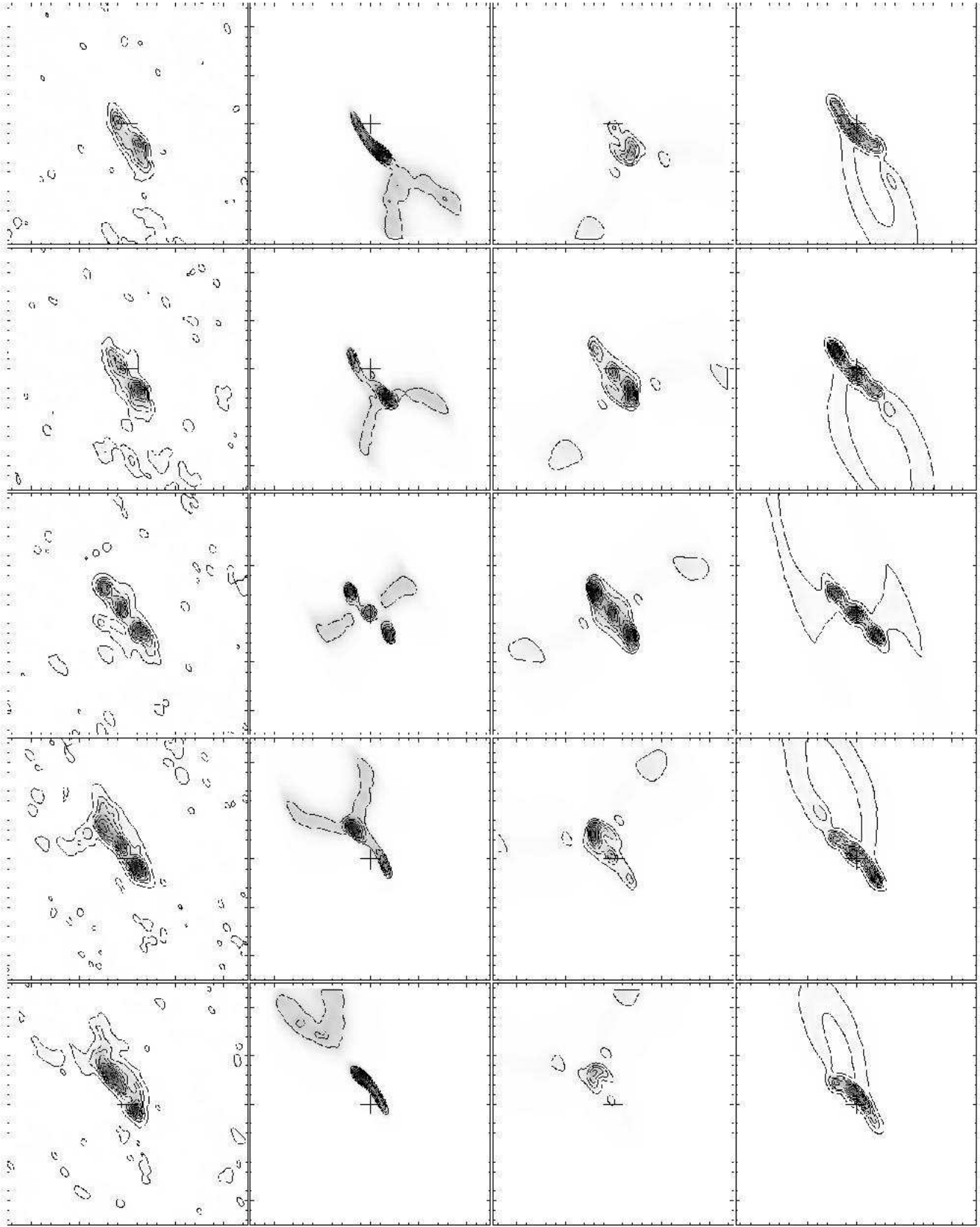}}
\end{center}
  \caption{Each row shows a 50$''$$\times$50$''$ channel map of the observed data, the axisymmetric model, the ellipse orbit model and the barred potential model respectively (from left to right) at a given velocity. From top to bottom the velocities are -80, -50, 0, 50 and 80\,km\,s$^{-1}$.  {We recall that, at 17\,Mpc, 1$''$ corresponds to 82\,pc along the major axis.}}\label{fig:modchannelmaps}
\end{figure*}
%
%

\begin{figure*}
\begin{center}
  
  \rotatebox{0}{\includegraphics[width=\textwidth]{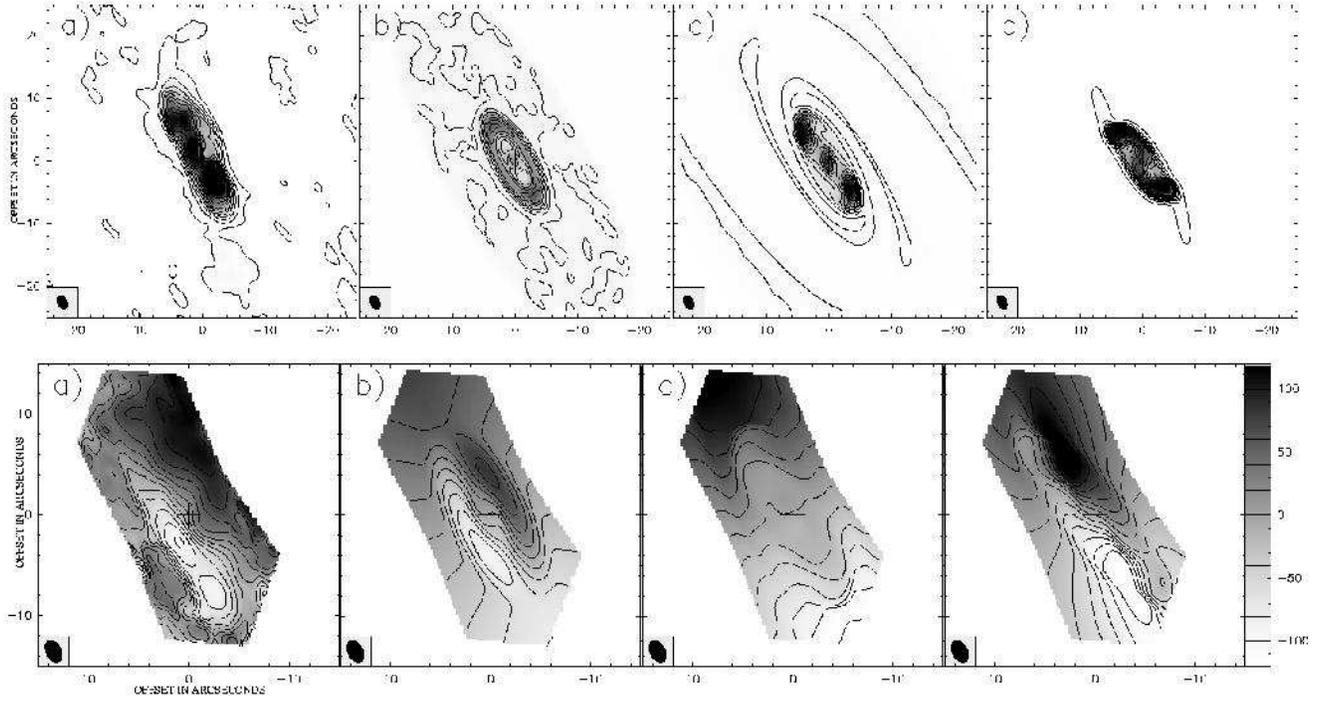}}
\end{center}
  \caption{Integrated maps (first row) and first moment maps (second row) for {\bf a)} the observed data, {\bf b)} the axisymmetric model, {\bf c)} the ellipse orbit model and {\bf d)} the barred potential model. In the observations, the velocity field can only be computed where  the signal-to-noise ratio is high enough, i.e. within a chevron shaped field. For comparison we used the same mask when computing the velocity fields from the models.}\label{fig:mean_vel}
\end{figure*}
\begin{figure*}
\begin{center}
 \rotatebox{0}{\includegraphics[width=\textwidth]{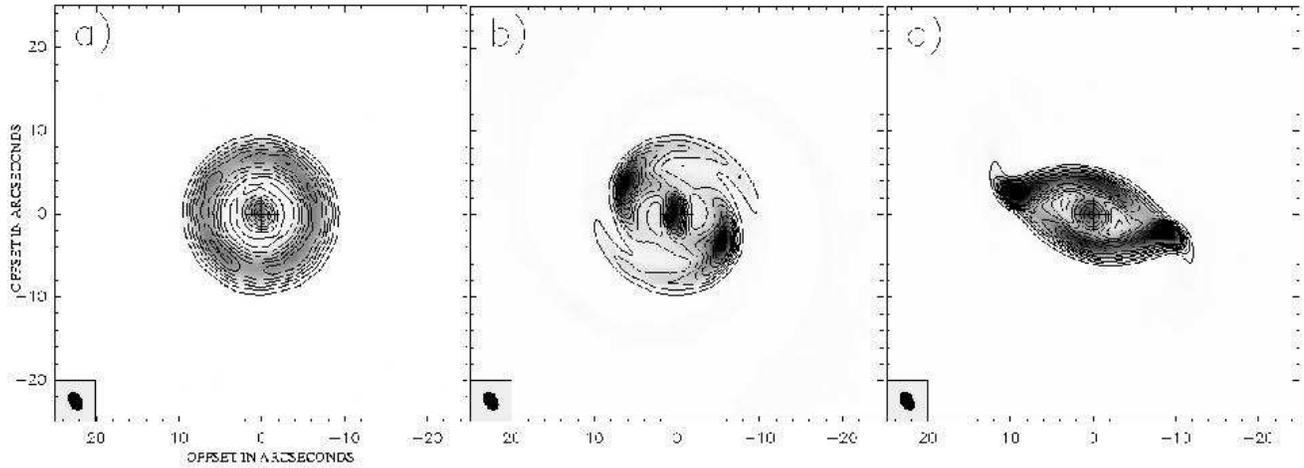}}
\end{center}
  \caption{Deprojected integrated CO(1--0) emission of {\bf a)} the axisymmetric model, {\bf b)} the ellipse orbit model and {\bf c)} the barred potential model. {We recall that, at 17\,Mpc, 1$''$$\sim$82\,pc along the major axis.}}\label{fig:mean_deproj}
\end{figure*}
\begin{figure*}
\begin{center}
 \rotatebox{0}{\includegraphics[width=\textwidth]{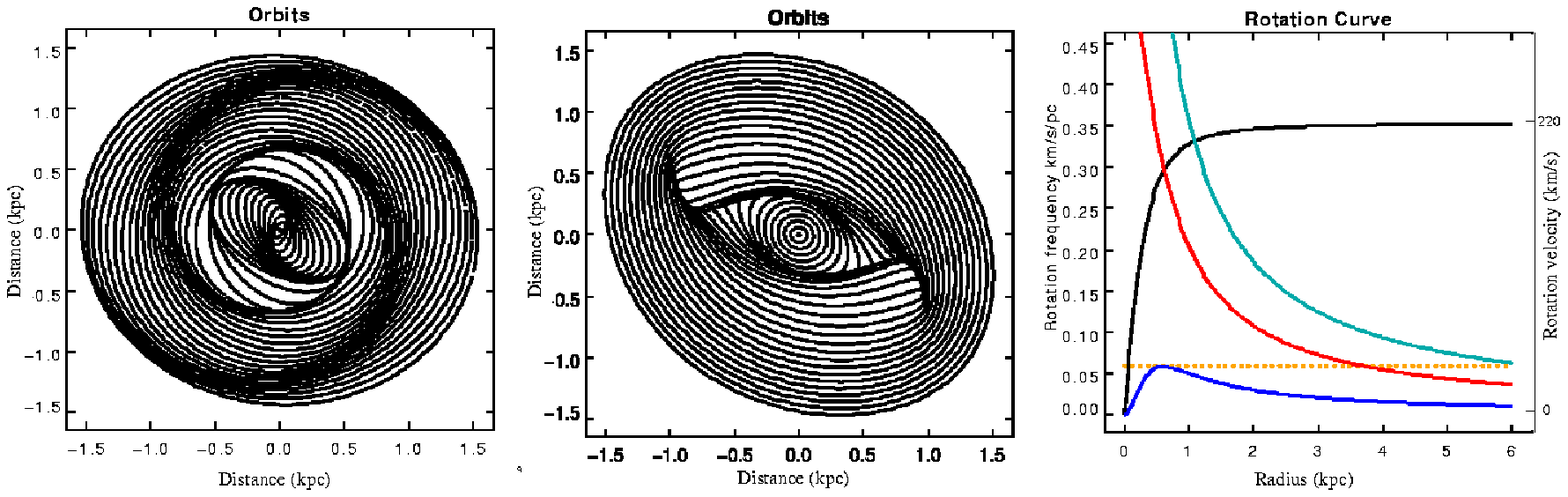}}
  
\end{center}
  \caption{Orbit patterns in the inner 1.5\,kpc of the elliptical orbit model (left) and the barred potential model (middle). Right-hand side: Resonance diagram, with the rotation frequency ($\Omega$) plotted in red, $\Omega-\kappa/2$ in blue, and  $\Omega+\kappa/2$ in cyan. The rotation curve is overplotted in black with the scale on the right-hand axis. }\label{fig:orbits}
\end{figure*}
\subsection{Motivation and strategy}
How can the CO emission described in the previous section be interpreted in terms of the radial distribution and kinematics of the molecular gas in the plane of the galaxy? The two outer peaks seen in the integrated map could be due to a ring  or to concentrations at the ends of a nuclear bar. The elongated shape seen in the channel maps could result from radial motions (streaming motions inside a density wave or due to shocks at the ends of a bar) or to elongated orbits. To discriminate among these possible interpretations and obtain a quantitative understanding of the deprojected gaseous disk, three different kinematic models are compared to the observations. The approach followed is progressive, starting with a pure geometrical model and proceeding to more 'physical' models, i.e. with more  physical underlying hypotheses used to compute the molecular gas motions. This approach allows us to constrain the physical properties of the galaxy step by step and thus to converge more efficiently to a reasonable solution. This is more efficient because the most physical model is very sensitive to  variations of the parameters and therefore more difficult to fit without a priori constraints.

\subsection{General description of the modeling}

The first model is based on circular orbits with additional radial motions, while the second model is constructed from elliptical orbits. 
These are purely geometrical hypotheses, and reality is expected to be more complex. However, the symmetry of the CO emission in the datacube (i.e. with respect to the dynamical center and the systemic velocity) suggests the gas flow is well organized and dominated by $m$$=$2 modes. Thus, if we assume a pure  $m$$=$2 symmetry, comparing the observations to the first model can help to quantify the amplitude of non-circular motions and estimate to what extent they can be reproduced by pure radial motions. Comparison to the second model allows us to evaluate the alternative interpretation of non-circular motions, namely the possibility that the clouds populate near-to-closed orbits that are elongated. The third model is based on analytical solutions for gas particle orbits in a barred potential.

The first two models are entirely specified by a radially varying set of properties similar to that of the widely used 'tilted ring model' \citep{1974ApJ...193..309R}. The values of the parameters are specified at some galactocentric radii and linearly interpolated between those radii. The number of radii where the parameter values are specified and their values can be chosen freely, giving us the ability to reproduce any radial distribution to a good approximation. Simulated data are produced by randomly assigning gas particles to the disk according to the distributions specified, projecting them in the datacube and convolving by the synthesized beam. As most of the emission comes from within the PdBI primary beam full width at half maximum ($43^{\prime\prime}$ at 115\,GHz), the effect of primary beam attenuation is neglected. For more details about the fitting procedure see Appendix\,\ref{ap:allmodels}.

\subsection{Axisymmetric model}
\begin{table*}
\caption{Circular model parameters}             
\label{tab:circmod}      
\centering                          
\begin{tabular}{l c c c c c c c}        
\hline\hline                 
 radius (pc) & 1.0 &  170 & 340 &  560 & 800 & 1700 & 2800 \\
\hline                        
 rotation vel. (km\,s$^{-1}$) & 40.0 &  30.0 &  40.0 &  50. & 70.0 & 100.0 & 110.0 \\
 radial vel. (km\,s$^{-1}$) & 0.0 & 0.0 & 0.0 & -120.0&  0.0 & 0.0 & 0.0 \\
 col. dens. (no units) & 6.0 &  1.0 &  0.5 & 4.0 &  0.4 &  0.4 & 0.4 \\
 velocity disp. (km\,s$^{-1}$) & 40.0 & 10.0 & 10.0 & 20.0 & 10.0 & 10.0 & 10.0 \\
 scale height (pc) & 20.0 & 15.0 & 10.0 & 10.0 & 10.0 & 10.0 & 10.0 \\
\hline                                   
\end{tabular}
\end{table*}

\subsubsection{Results}
The parameters of the axisymmetric model (see details in Appendix\,\ref{ap:model1}) reproducing the main features of the observations with 7 radii are listed in Table\,\ref{tab:circmod}. The corresponding channel maps for 5 velocities are shown in the second column of Fig.\,\ref{fig:modchannelmaps}, while the corresponding integrated intensity and velocity maps are shown in Fig.\,\ref{fig:mean_vel}\,b. A face-on view of the model is shown in Fig.\,\ref{fig:mean_deproj}\,a.

The three peaks in the integrated map are reproduced by a ring with mean galactocentric radius of $\sim$600\,pc and a central concentration. The column density must be higher by a factor of $\sim$10 inside the ring (at 600\,pc) with respect to the density at 300\,pc and 700\,pc. The column density must increase again inside a radius of 300\,pc to reach a value higher by a factor of $\sim$15 at the center. The distribution of the CO emission in the channel maps requires an inward radial velocity of 120\,km\,s$^{-1}$ in the ring at 600\,pc. Lower radial velocities in the ring fail to reproduce the data.

This model can reproduce the shape of the CO emission in the channel maps at all velocities. The slight shift of the emission toward the northwest from channel to channel and, as a consequence, the twisted isovelocity lines are also well reproduced (compare Fig.\,\ref{fig:mean_vel}a with \ref{fig:mean_vel}b). Moreover, the ring interpretation provides a natural explanation for the fact that the elongated shape is aligned with the major axis of the galaxy: it is merely an inclination effect. The width of the emission along the minor axis, however, is  larger than in the observations, and the two symmetrical peaks are not as highly contrasted as observed. 
Also, the radial velocity in the ring required to reproduce the observations is extremely high and cannot be interpreted as the manifestation of real radial motions. 
\subsubsection{Interpretation}
We conclude from this first modeling that, although a ring of 600\,pc radius with a central concentration can explain to some extent the spatial distribution of the CO emission,  the amplitude of the non-circular motions is unrealistically high, i.e. comparable to the amplitude of the rotation motions. No physics could explain the collapse of a ring at this speed . This result implies that pure radial motions alone cannot explain the departures from circular rotation, which must be at least partly due to the shapes of the orbits themselves. Paradoxically, while the best fit solution of the circular model reproduces the observations well, it implies that the model is not a good representation of reality and that elongated orbits must be considered.

\subsection{Elliptical orbit model} \label{sec:ellmodel}
\subsubsection{Description}
\begin{table*}
\caption{Elliptical orbit model parameters}             
\label{tab:ellmod}      
\centering                          
\begin{tabular}{l c c c c c c c c c c}        
\hline\hline                 
charact. radius (pc) & 1.0 &  85& 170 & 510 &  620 & 680 & 850 & 1700 & 2800 & 5500 \\
\hline                        
 charact. rot. vel. (km\,s$^{-1}$) & 40 &  25 &  10 &  20 & 40 & 90 & 120 & 150 & 200 & 220 \\
ellipticity & 0.8 & 0.8 & 0.8 & 0.7 & 0.5 & 0.3 & 0.3 & 0.3 & 0.2 & 0.2\\
position angle (deg) & 110 & 130 & 160 & 210 & 160 & 100 & 20 & -180 & -230 & -330 \\
 col. dens. (no units) & 3.0 &  2.0 &  0.2 & 0.7 &  0.8 &  0.3 & 0.03 & 0.03 & 0.03 & 0.02 \\
 velocity disp. (km\,s$^{-1}$) & 30 & 35 & 40 & 40 & 40 & 30 & 10 & 10 & 10 & 10  \\
 scale height (pc) & 20 & 18 & 17 & 15 & 10 & 10 & 10 & 10 & 10 & 10 \\
\hline                                   
\end{tabular}
\end{table*}
Although elongated orbits are not expected to be  elliptical in general, ellipses provide a first order approximation to any non-circular closed or nearly-closed shape. To estimate a possible contribution from the orbit elongation to the non-circular motions {and to get a picture of the orientations of the orbits in the galactic plane (the orbit pattern)} we fit a model in which the gas particles populate elliptical orbits and their angular momentum is constant over the orbit. ``Nearly-closed'' orbits means orbits that are stable on the time scale of a rotation period, which is possible only if the orbits do not intersect, since otherwise the cloud shocks would re-orient the gas on different orbits. Thus, for this model to be self-consistent, the variations in position angle and ellipticity must satisfy this 'no crossing' condition.
The density of clouds on a given orbit is not uniform. If it were assumed independent of the rotation sense it would be inversely proportional to the velocity. However, it is known that molecular clouds are formed in density waves or shocks, from which stars form that then destroy the parent cloud. The density is therefore expected to increase abruptly when entering the density wave or shock and to decrease smoothly after. In this simple model, density waves correspond to overdensities due to orbit crowding, i.e. are kinematic waves: they form at the aphelia of the ellipses when the position angles change with radius. To mimic the formation/destruction of the clouds  we  apply an exponential taper to the density. The application of this taper significantly improves the model fit. More details about this model are given in Appendix\,\ref{ap:model2}.
\subsubsection{Results}
The parameters of such a model {best reproducing the observations} with 10 radii are listed in Table\,\ref{tab:ellmod}. The corresponding channel maps for 5 velocities are shown in the third column of  Fig.\,\ref{fig:modchannelmaps}, while the corresponding integrated map and velocity field are shown  in Fig.\,\ref{fig:mean_vel}\,c. The face-on view of the model solution is shown in Fig.\,\ref{fig:mean_deproj}\,b and the orbital pattern in Fig.\,\ref{fig:orbits}\,a.
Strong variations in the column density,  the angular momentum,  the orbit orientations and the velocity dispersion are required at a radius of  $\sim$600\,pc to reproduce the peaks in the integrated map and the shape of the distribution in the channel maps. As in the previous model the column density must increase by an order of magnitude at  $\sim$600\,pc, however inside the $\sim$600\,pc radius the column density only drops by a factor of 4. A central concentration is still required within the innermost 150\,pc. The angular momentum must decrease  abruptly inside  $\sim$600\,pc.
The velocity dispersion must be  $\sim$40\,km\,s$^{-1}$ at $\sim$600\,pc. {The most remarkable result is that to reproduce the elongated core in the channel maps, it is necessary to invert the variation of the orbit position angles at the same radius of the column density peak (in the range 500--600\,pc). The orbit position angle must be maximum at this radius and equal to $\sim$$30 \bmod 180\deg$. This feature of the orbit pattern seems to be strongly constrained by the observations.} 

From  Fig.\,\ref{fig:modchannelmaps} it can be seen that the low level emission is well reproduced by this model, giving us confidence that this emission is real and that the rotation velocity of the disk at 5.5\,kpc is $\sim$220  \,km\,s$^{-1}$. This value is consistent with the estimates of \citet{1988AJ.....96..851G} based on H\,{\sc i} observations and  \citet{1999AJ....118..236R} based on optical observations. 
%
%

The main differences between the observed and the modeled channel maps occur at high velocities, with the model being unable to reproduce the emission closest to the center seen up to $\pm$150\,km\,s$^{-1}$. Moreover the model cannot reproduce the continuous shift of the emission to the northwest and therefore the central velocity gradient present in the velocity field.

In addition, to obtain emission that is spatially elongated up to high velocities, we must set a velocity dispersion of at least 40\,km\,s$^{-1}$ in the ring. This is  a high velocity dispersion, a fraction of this may come from a shock, but beam averaging of the velocities at the ends of the orbits can also be expected to contribute significantly. This consideration implies that the real orbits must have sharper ends than ellipses.
\subsubsection{Interpretation}
The orbit pattern and the face-on view of the emission strongly resemble gas kinematics in a barred potential as observed in many galaxies. The large gradient of the orbit position angle at 600\,pc mimics a tightly wounded spiral forming a ring. {The change in  sign of the gradient of the position angle is a remarkable feature that is known to occur at the Inner Lindblad Resonance (ILR) when there is only one, or between ILRs when there are two of them \citep[see e.g. ][]{1994PASJ...46..165W}. The fact that this feature coincides with a column density peak, which is expected at resonances, indicates that this is actually a single ILR}.
The two symmetric peaks in the face-on integrated map and the large velocity dispersion at the resonance would then correspond to the shocks frequently observed  at the locations of dust lanes in bars, which are due to the crowding of orbits that favors cloud collisions \citep[see e.g.][]{1996ASPC...91..286C}. 
Identification of the ILR is further supported by the orientation of the bar inferred from the model, which is similar to that of the large scale stellar bar. 
Indeed, the shocks are expected to lead the bar \citep[see e.g.][]{1996ASPC...91..286C}, which would therefore have a position angle somewhere between 0 and 30\,$\deg$ (see Fig.\,\ref{fig:mean_deproj} and see also the maximum value for the orbit position angle in  Table\,\ref{tab:ellmod}: 210\,$\deg$$=$$30 \bmod 180\deg$). This is consistent with the large scale stellar bar position angle of 15\,$\deg$ \citep{2005ApJ...630..837J}. This agreement implies that the resonance observed is unlikely to result from the action of an inner bar, but is  rather the ILR of the large scale bar.

Is this interpretation supported by a model based on gas orbits in a barred potential? Why is the velocity field with the strong velocity gradient twist in the center not reproduced? Is this due to the elliptical approximation of the orbits or are pure radial motions (as opposed to radial motions intrinsic to non-circular but closed orbits) also contributing? To answer these questions, we compare the observations to a model based on analytical solutions for the gas orbits in a barred potential.
\subsection{Barred potential model}\label{sec:barmodel}

\begin{table*}
\begin{minipage}[t]{\columnwidth}
\caption{Barred potential model parameters}             
\label{tab:barmod}      
\centering                          
\renewcommand{\footnoterule}{}  
\begin{tabular}{l c c c c c c c c c c}        
\hline\hline 
galaxy position angle (deg) & 20 \\
characteristic length (pc) & 400$^{\mathrm{a}}$\\ 
characteristic velocity (km\,s$^{-1}$) & 220$^{\mathrm{a}}$\\ 
bar position angle (deg) & 20 \\
bar strength & 0.015 \\
pattern speed (km\,s$^{-1}$\,kpc$^{-1}$) & 60$^{\mathrm{a}}$\\ 
radial dissipation rate (km\,s$^{-1}$\,pc$^{-1}$) & 0.002 \\
azimuthal dissipation rate (km\,s$^{-1}$\,pc$^{-1}$) & 0.005 \\
\hline\hline                 
charact. radius (pc) & 10  & 250 & 400 & 600 & 700 & 1500 & 3000 \\
\hline  
col. dens. (no units) & 1.5 & 0.1 & 0.5 & 1.0 & 0.05 & 0.02 & 0.02 \\
velocity disp. (km\,s$^{-1}$) & 35 & 10 & 30 & 30 & 10 & 10 & 10 \\
scale height (pc) & 10 & 10 & 10 & 10 & 10 & 10 & 10\\
\hline                                   
\end{tabular}
\end{minipage}
\begin{list}{}{}
\item[$^{\mathrm{a}}$] value fixed by the elliptical orbit model solution
\end{list}
\end{table*}
\subsubsection{Description}
This model was introduced and developed by \citet{1975BAAS....7Q.513S}, \citet{1994ASPC...66...29L}, \citet{1994PASJ...46..165W}, \citet{1999ApJS..124..403S} and \citet{2000PhDT.........6B}. It is based on analytical solutions for the orbits of gas particles obtained by a perturbation calculation of the epicyclic stellar motions.
The potential is described by an axisymmetric component and a weak $m$$=$2 perturbation. The perturbation has the same radial profile as the axisymmetric component and is modulated by a sine in azimuth. The amplitude of the perturbation relative to the axisymmetric component is defined as the strength of the bar. {For the radial profile of the potential we use the logarithmic shape. This choice is justified by the ratio of the resonance radii. Indeed, the corotation radius of the large scale bar,  assumed to be 1.2 times the bar extension, is  $\sim$5\,kpc according to \citet{2002MNRAS.337.1118L} and according to the elliptical orbit model the ILR radius is 600\,pc, implying  $R_{\rm c}/R_{\rm ILR}$$\sim$8. As shown in Appendix\,\ref{app:resonances}, for a logarithmic potential this ratio is 6.3. For comparison a Plummer potential gives a ratio of $\sim$3 \citep{1994PASJ...46..165W}.}

As in \citet{2000PhDT.........6B}, two dissipation terms are introduced that characterize the damping of radial and azimuthal oscillations and are assumed to reproduce the dissipative behavior of gas particles. These terms cause the major axis of the orbits to lead the bar, the offset being maximal at the ILR. The limitations of the model are (1) it is singular at corotation; (2) the epicyclic motions used to compute the perturbations are based on the axisymmetric component only so that the model is not fully self-consistent; and (3) computing closed orbits assuming dissipation rates is not self-consistent, because dissipative particles imply open orbits. As a consequence, the radii of the resonances obtained with this model (Appendix\,\ref{app:resonances}) cannot be taken as accurate results.  However, this can be seen as a first order approximation that works only when the real orbits are near to being closed. In this case the resonance radii obtained give a basic picture of the scales involved. The orbit patterns produced by this model exhibit some of the main features expected in barred potentials, e.g., the elongated orbits, the phase shift change at an ILR, and the spiral pattern. Also, they match observations and results of numerical simulations \citep[see][]{1994ASPC...66...29L}. This analytical approach should be particularly suitable for  NGC\,4569 which, as noted in Section\,\ref{sec:obs}, is rather symmetric and dominated by an $m$$=$2 component in CO and  shows a weak stellar bar in the NIR \citep[][]{2002MNRAS.337.1118L}. The results of the elliptical orbit fit allow us to fix some of the parameters describing the potential.
More details about this model are given in Appendix\,\ref{ap:model3}.

\subsubsection{Results}
The best fit parameters for 7 radii, including the parameters fixed by the results of the elliptical orbit fit, are listed in Table\,\ref{tab:barmod}. Some of the channel maps of the simulated data are shown in the last column of Fig.\,\ref{fig:modchannelmaps}; the  integrated map and velocity field are shown in Fig.\,\ref{fig:mean_vel}\,d. The face-on view of the model solution is shown in Fig.\,\ref{fig:mean_deproj}\,c and the orbital pattern in Fig.\,\ref{fig:orbits}.
The best fit bar position angle of 20\,$\deg$ is consistent with the large scale stellar position angle  of 15\,$\deg$ derived by \citet{2005ApJ...630..837J}. This result confirms the conclusion drawn from the elliptical orbit model, namely that the kinematics inside the inner kiloparsec are driven by the gravitational potential of the large scale stellar bar\footnote{Note that although this conclusion drawn from the elliptical orbit model was used as an hypothesis to fix some of the parameters of the model, these fixed parameters are not related to the bar position angle, so the argument is not circular.}.

For this model as for the others, a column density jump of an order of magnitude is required at a radius of 600\,pc and a central concentration of similar column density is required.  A velocity dispersion of 30\,km\,s$^{-1}$ is required at the radius of the column density increase. Although this is still a high value, it is less than for the elliptical orbit model because the orbits have much sharper ends  (Fig.\,\ref{fig:orbits}) that  contribute significantly to the velocity dispersion through beam averaging. This suggests that the orbit shapes in this model are more realistic than the ellipses.

As can be seen in  Fig.\,\ref{fig:modchannelmaps}, this model reproduces the data better than the elliptical orbit model, especially at high velocities. The low level emission is not so well reproduced but this is expected as this model diverges at corotation,  making it unreliable at large radii. 
The velocity field  (Fig.\,\ref{fig:mean_vel}) is also the closest to the data when compared with the two other models discussed above.
\subsubsection{Interpretation}
The fact that an analytical model that includes certain physical assumptions (about the potential shape and gas properties) reproduces the data well lends support to its underlying assumptions. In particular, the interpretation of the column density increase at 600\,pc as an ILR of the large scale bar is well confirmed by this fit. This also supports the assumption that there is only one ILR or that the ILR region is small with little space for x$_2$ orbits. If there were two ILRs, the gradient of the orbit position angle would change in sign at a similar radius \citep[in the middle of the 2 ILRs as described in ][]{1994PASJ...46..165W}, but the maximum position angle of the orbits would be much greater, leading to a much greater position angle for the oval ring unlikely to fit the observations equally well.
The good fit also shows that even though this model is not fully self-consistent, the orbit pattern and orbit shapes favored by its best-fitting version may be reasonably close to reality. {We note that the configuration of the orbit pattern implies that the velocity field is very sensitive to the orbit shapes. Indeed, the most elongated orbit are almost aligned with the major axis (the bar is almost aligned with the major axis), and the strongest velocity components are therefore orthogonal to the line of sight. As a consequence the velocity field is not only dominated by the elongation of the orbits but it is also sensitive to the orbit shapes, i.e. how the velocity vectors behave at the ends of the orbits. This can explain why the diffence between the ellipses and these more realistic orbits implies a significant difference in the velocity field. }
\section{Mass inflow}\label{sec:massinflow}
\begin{figure*}
\begin{center}
\rotatebox{0}{\includegraphics[width=\textwidth]{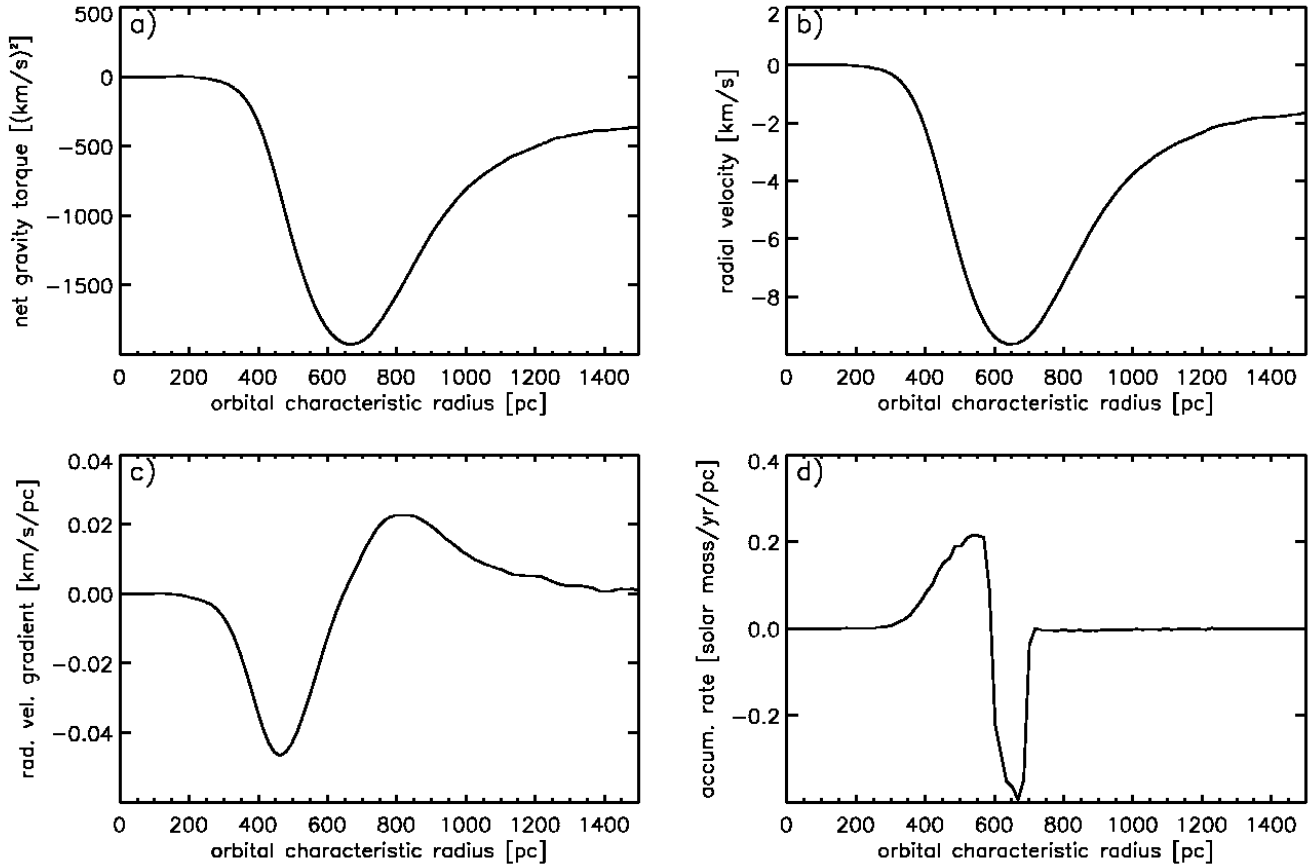}}
 
\end{center}
  \caption{{\bf a)} Net gravity torque,  {\bf b)} radial velocity, {\bf c)} radial velocity gradient and  {\bf d)} mass accumulation rate versus characteristic radius of the orbits (see Section\,\ref{sec:massinflow} for details). All the quantities computed for the maximum bar strength, $\epsilon$=0.05.}\label{fig:torque}
\end{figure*}
 Two mechanisms contribute to angular momentum dissipation and mass inflow: gravitational torques, which depend on the potential and on the individual gas orbit shapes and orientations, and viscosity torques, which depend on the cloud-cloud collision rate and on the interstellar turbulence.
In the previous section the observed molecular gas kinematics were successfully modeled by analytical solutions for gas particle orbits in a barred potential. We now address the question of the mass inflow based on this model solution.

As the potential of the model solution is known at each point in the galaxy, the gravitational torque can be directly computed for each particle following Eq.\,\ref{eq:insttorque}. 
 In addition, as the orbits are known, the torques can be integrated over the orbits to estimate the angular momentum loss over a rotation (Eq.\,\ref{eq:angmomloss}) as well as the corresponding radial velocity of the matter inflow (Eq.\,\ref{eq:radvel}).  As already outlined, this approach is not fully self-consistent, since the orbits should not be closed if the angular momentum changes over a rotation, but the angular momentum loss thus computed can be considered as a first order approximation to the case where orbits are nearly closed, i.e., the angular momentum loss due to the gravitational torque is a small fraction of the angular momentum. The good fit of the model to the observations suggests that, on average, the molecular gas in the inner kiloparsec lies in this regime.

The torque is proportional to the bar strength, which is only poorly constrained by the 'weak bar' regime requirement (see discussion above), to $\epsilon$$<$$0.05$. This implies that only upper limits can be derived for the mass inflow. Figure\,\ref{fig:torque} shows the net torque (as defined Eq.\,\ref{eq:insttorque}) for $\epsilon$$=$0.05  on the top left, and the corresponding radial velocity on the top right. Both curves look similar, because the radial change of the angular momentum as determined by the axisymmetric potential is almost constant above 500\,pc (see Eq.\,\ref{eq:radvel}).
Due to their circular shape, the orbits with a characteristic radius less than 300\,pc undergo no net torque and as a consequence the large scale bar would not be able to fuel the nuclear starburst on these scales via gravitational torques only. The maximum radial velocity is of order 10\,km\,s$^{-1}$ and occurs for the orbit with a characteristic radius of 650\,pc, i.e. at the ILR. The value of this upper limit for the radial velocities  is consistent with our working hypothesis, namely, that to first order, radial motions can be neglected compared to rotational motions, consistent with the results of \citet{2004ApJ...605..183W}. The variations of the radial velocity, shown  on the bottom left of Fig.\,\ref{fig:torque}, imply that the radial flow is not steady over the disk and as a consequence that matter must accumulate in some orbits and be depleted in others. The actual mass flow also depends on the gas mass available at each radius (the radial velocity would directly translate into mass flow only if the gas mass did not depend on the radius, i.e., for a column density profile following $R^{-1}$). Nevertheless, without any assumption about the radial density profile, the variation of the radial velocity shows that the gas will tend to be depleted just outside the ILR (the inward velocity increases with decreasing radius down to 600\,pc) and to accumulate inside the ILR (the inward velocity decreases with decreasing radius inside 600\,pc). This model is therefore consistent with the formation of rings at ILR as observed in many spiral galaxies \citep[][]{1996FCPh...17...95B} and often used as a signature of the ILR. 

The mass accumulation rate per radial element at a given radius is the gradient of the  radial momentum (product of the radial linear density with the radial velocity); it is plotted as a function of the characteristic radius for $\epsilon$$=$0.05  in  Fig.\,\ref{fig:torque}\,d. To obtain the correct mass units, the column density profile of the model was normalized so that the total mass of gas equals that measured in Section\,\ref{sec:obs}. We see that the maximum accumulation rate inside the ILR is of order 0.2\,$M_{\odot}$\,yr$^{-1}$\,pc$^{-1}$. Outside and close to the ILR the depletion rate is of similar amplitude. The accumulation rate integrated inside the ILR (i.e. over the radius range 300--600\,pc) gives an inflow rate of order  20\,$M_{\odot}$\,yr$^{-1}$. This is significantly higher than other inflow rates estimates in the literature which are in the range 1--4 \,$M_{\odot}$\,yr$^{-1}$ \citep{1995ApJ...441..549Q, 1999ApJ...511..709L, 1997ApJ...482L.143R}. {But this is an upper limit for this specific model and lower values are not excluded}. The accumulation of molecular gas inevitably leads to a density increase and star formation. If the fraction of infalling gas transformed into stars (star formation efficiency) is constant and is of order 10$\%$, then the star formation rate (SFR) integrated over the accumulation region would be of the order of a few  solar masses per year, which is realistic when compared to what is observed at ILRs in other galaxies. In NGC\,4569 however, the H${\alpha}$ morphology and the CO(2--1)/(1--0) line ratio suggest that most of the star formation observed in this galaxy is taking place inside the inner 100\,pc (see Section\,\ref{sec:optical}). {The presence of $\sim$5$\times 10^{4}$ O and B stars in the inner 30\,pc \citep{2002AJ....124..737G}  would imply a constant SFR of order 2\,$M_{\odot}$\,yr$^{-1}$ (assuming a Salpeter IMF and a life time of 5\,Myr for O and B stars).
This implies either that another fueling mechanism is taking over inside the inner  300\,pc, or that the nuclear starburst is ending and star formation at the ILR will dominate in the future.}

Viscosity torques are intrinsic to  the model and are specified through the radial and azimuthal dissipative parameters. However, as outlined in \citet{2000PhDT.........6B}, these parameters lead to closed orbit solutions, so they cannot be directly used to compute the dissipation of angular momentum and the mass inflow rate. 
Estimating the intensity of the viscosity torques based on the density and velocity dispersion at a given point in the disk would require adopting  prescriptions for the visosity that are generally acknowledged as very uncertain. Thus, our model does not allow us to directly constrain the rate of mass inflow due to viscosity torques. It could well be that these torques take over the fueling of the nucleus inside the 300\,pc where the gravity torques are not efficient anymore. Alternatively, an inner stellar bar could drive the gas further inward as in NGC\,6946 \citep{2006ApJ...649..181S}; however, the current observations do not show any evidence of such a small inner bar.

\section{Discussion}\label{sec:discussion}
Models of gas dynamics in galaxies come in two main categories: ballistic models in which the gas is represented by colliding clouds on ballistic orbits \citep{1985A&A...150..327C, 1994AJ....108..476B}, and hydrodynamic models in which the gas is assumed to be a continuous dissipative medium \citep{1992MNRAS.259..345A,1995ApJ...449..508P,1995MNRAS.277..433W, 1993A&A...268...65F}. Comparisons of the two approaches are discussed in \citet{1994ASPC...66...29L}, \citet{1994PASJ...46..165W},  \citet{1996ASPC...91..286C} and  \citet{1999ApJ...526...97R}. The analytical model of orbits in a barred potential presented in the previous section effectively traces the evolution of gas particles in the gravitational potential as they are subjected to collisions. Although the equations are continuous, they are not based on hydrodynamic assumptions, and the gas is assumed to be an ensemble of independent particles rather than a continuous medium, meaning the model can be classified as ballistic. The good fit of this model to our CO observations presented here seems to validate such a ballistic approach. The results of the two approaches may actually not have so many differences that the observations can distinguish. As already suggested by \citet{1994ASPC...66...29L}, it may be that the orbits out of this analytical model describe a gas flow similar to the one obtained with a hydrodynamic model. This hypothesis is supported by comparing the orbit pattern as shown in Fig.\,\ref{fig:orbits}\,b to the gas flow resulting from a hydrodynamic simulation as shown in Fig.\,9 of  \citet{1999ApJ...526...97R}. The shapes of the orbits resemble the gas streamlines in their model and the dust lane/shock region of their model is well reproduced by the orbit crowding in our model. The main difference is that the largest circular orbit in our model is located well inside the ILR, whereas it defines the ILR in their model. As a consequence, the ring formed at the ILR by gas accumulation is elongated in our model (Fig.\,\ref{fig:mean_deproj}\,c) rather than circular. But such a shape has also been seen  by \citet{2005PASJ...57..905N} based on  Smoothed Particle Hydrodynamics (SPH) simulations of the same galaxy. 

The difference between ballistic and hydrodynamic models outlined in  \citet{1999ApJ...526...97R} corresponds to the sharp inversion in the radial component of the velocity vector in the shock/dust lane region. This feature is argued to be reproduced by the hydrodynamic models but not by the orbits of ballistic models.  However, this argument is based on the assumption that the orbits are elliptical (see their Fig.\,10). As emphasized in Section\,\ref{sec:barmodel} the analytical orbits of gas clouds in a barred potential used here have much sharper ends than ellipses, blurring the nominal distinction between hydrodynamic and ballistic models.

It is also interesting to note that the conclusions of both approaches in terms of mass inflow are not so different. The inflow stops at the largest circular orbits (the ``nuclear ring'') in the model of  \citet{1999ApJ...526...97R} and at the largest x$_2$ orbit or at the largest nonlooping x$_1$ orbit in \citet{2004ApJS..154..204R}. In all cases, the bars seem to be incapable of driving molecular gas inside a radius of 100\,pc. Similarly, in our model, the gravitational torques stops where the orbits become circular at a radius of 300\,pc. 

The advantage of the method followed here is that we are able to first identify the ILR based on a kinematic argument, namely the inversion of the radial variation of the ellipse orientations in the second model (Section\,\ref{sec:ellmodel}). This allows us to fix the potential parameters of the third model, i.e. the orbit pattern, and to fit the radial profile of the column density and velocity dispersion independently. Although limited in its domain of application (weak bars and inside corotation), this method has the advantage of being insensitive to initial conditions or numerical effects that may affect evolutionary models (ballistic or hydrodynamic). In addition, it is obviously very undemanding in terms of computation resources and well adapted to iterative fitting to observations.
\section{Conclusion}\label{sec:conclusion}

The main results of this article can be summarized as follows:
\begin{enumerate}
\item In NGC\,4569, 70$\%$ of the $[1.1\pm0.2]\times10^9$\,M$_{\odot}$ of molecular gas detected in the inner 20$''$ is concentrated within a radius of 800\,pc and distributed along the large scale bar, with two peaks at $\sim$500\,pc from the center and one peak close to the center. There is a hole in the CO distribution at the location of the nuclear starburst where evidence of outflow is seen in H$\alpha$ emission.
\item {Fitting a model based on elliptical orbits reveals a change in sign of the radial variation of the orbit orientations at a radius of $\sim$600\,pc, where molecular gas is also accumulated.  This coincidence,  together with  the value of the orbit position angle maximum, leads us to identify this feature with confidence as a single ILR of the large scale bar.}  
\item The kinematics are well reproduced by a model based on analytical solutions for the gas particle orbits in a weakly barred potential {with a single ILR at 600\,pc.}
\item The gravitational torques implied by this model are able to efficiently funnel the gas inside the ILR down to a radius of 300\,pc but another mechanism must take over to fuel the nuclear starburst, i.e., there is no evidence for ongoing fueling of the starburst by the large scale bar. If such a mechanism does not exist, the nuclear starburst will stop and  star formation at the ILR will dominate with a star formation rate of order one solar mass per year (assuming a 10$\%$ star formation efficiency). 

\end{enumerate}

This last point agrees with the result of \citet{2005A&A...441.1011G}, namely that the gravitational torque caused by kiloparsec scale perturbations of the potential  seems not to be the only mechanism responsible for fueling galactic nuclei.

\appendix
\section{Detailed description of the modeling}

\subsection{All models}\label{ap:allmodels}

The first two models are entirely specified by a radially varying set of properties similar to that of the widely used 'tilted ring model' \citep{1974ApJ...193..309R}. The values of the parameters are specified at some galactocentric radii and linearly interpolated between those radii. The third model has additional global parameters.
Thus, the total number of free parameters is adjustable through the number of radii, $N_{\rm radii}$, and is given by $N_{\rm radii}\times N_{\rm properties}+N_{\rm global}$, where  $N_{\rm properties}$ is the number of properties specified at each radius including the value of the radius. The number of radii required to reproduce the data satisfactorily was found to be greater than 7. In this case, for all models the number of free parameters is too high compared to the dynamic range of the data (signal-to-noise ratio, spatial dynamic range and spectral dynamic range) to permit a rigorous minimization of an objective function like $\chi^2$ over the full parameter space.
Since the goal of this modeling is to test the compatibility of the different hypotheses with the main features of the data, rather than to estimate to a high accuracy the numerous free parameters, the comparison of the simulated data to the real data has been done by eye\footnote{This comparison uses the DALIA interface to numerical simulation codes which is being developed jointly by the Observatoire de Paris and the Max Planck Institut F\"ur Radioastronomie \citep{2006EAS....18..299B}. This graphical user interface gives access to the model parameters and allows the user to directly compare the outputs of the code with real observations.}.

{In all the models, the inclination of the galaxy is fixed to 70\,$\deg$ \citep{1988ngc..book.....T}, its position angle to 30\,$\deg$ \citep[][ except in the third model where the PA is also a free parameter]{1988AJ.....96..851G} and its distance to 17\,Mpc \citep[][ only used to convert angular scales into physical scales]{1983A&A...118....4B, 1988ngc..book.....T}}. In addition, in all models we make the effective assumption that mass column density is proportional to the CO intensity with the same constant of proportionality everywhere.
\subsection{Axisymmetric model}\label{ap:model1}
%
%
In the first model, the properties specified are the radii, the rotational velocity, the radial velocity (positive in outward direction), the column density through the disk, the disk scale height (FWHM of the Gaussian profile) and the local velocity dispersion (square root of the Gaussian distribution variance). The number of free parameters is therefore $6\times N_{\rm radii}$. The vertical distribution of matter and the velocity distribution at a given position are assumed to be Gaussian. 
\subsection{Elliptical orbit model}\label{ap:model2}
To mimic the formation/destruction of the clouds  we  apply an exponential taper to the column density along the orbits.
The taper equals 1 at the aphelion and decreases like $\exp(-2\theta /\pi)$, where $\theta$ is the azimuth along the ellipse. The sign of $\theta$  is determined by the sense of rotation of the cloud in the reference frame of the density wave for which a pattern speed needs to be specified. Here we assume that all the gas lies within the corotation radius, and that the density decreases smoothly in the sense of rotation of the whole galaxy, i.e. counter-clockwise.

The properties of the model are the characteristic radius, the characteristic rotation velocity (the angular momentum), the ellipticity, the position angle of the ellipse, the column density, the velocity dipersion and the scale height. The pattern speed of the entire model was not well constrained by the observations and since the larger the value the worse the fit, it was decided to fix this parameter at zero.
 The number of free parameters is therefore $7\times N_{\rm radii}$. As in the previous model, the vertical distribution of matter and the velocity distribution at a given position are assumed to be Gaussian. 

\subsection{Barred potential model}\label{ap:model3}

We use 2 parameters describing the logarithmic potential (the characteristic scale and velocity, {see Eq.\,\ref{eq:potential})}, 3 parameters for the bar perturbation (the position angle relative to the galaxy major axis, the strength, and the pattern speed) and 2 parameters for the gas dissipative properties (the radial and azimuthal dissipation rates).  We found it difficult to obtain a good solution with a galaxy position angle of 30\,$\deg$ as in the previous models, so we left this as an additional free parameter here. In addition, the gaseous disk is described by 3 properties (the column density, the velocity dispersion, and the scale height) at different radii as in the previous models (the radius here is the characteristic radius of the orbit, which corresponds to the radius of the circular orbit without bar perturbation). The number of free parameters is therefore $8+4\times N_{\rm radii}$. 

The results of the elliptical orbit fit allow us to fix some of the parameters describing the potential. First, the characteristic velocity of the axisymmetric component of the potential corresponds to the rotation velocity at infinity. Based on previous results, we take $v_{\rm p}=220$\,km\,s$^{-1}$, also consistent with the value of \citet{1988AJ.....96..851G} based on H\,{\sc i} observations.  Next, the fact that the ILR was identified at 600\,pc from the center in the elliptical orbit model allows us to fix the characteristic scale of the potential (using Eq.\,\ref{eq:ri}) to $r_{\rm p}=400$\,pc. Finally, assuming there is only one ILR (as the elliptical orbit model suggests), the pattern speed is set by Eq.\,\ref{eq:pat} to $\Omega_{\rm p}$=$60$\,km\,s$^{-1}$\,kpc$^{-1}$. The remaining global parameters (as opposed to those specified at each radius) are the bar strength, its position angle and the 2 dissipation rates. There is a balance between the bar strength and the dissipation rates, in the sense that the stronger the bar or the smaller the dissipation rates the closer the orbits are to pure stellar orbits. Stellar orbits are organized in families, the so-called x$_1$ and x$_2$ orbits, \citep{1989A&ARv...1..261C, 1993RPPh...56..173S}, which are orthogonal to each other, intersect, and  can exhibit large loops. As long-lived gas orbits cannot intersect or have loops, a realistic solution for the orbits must keep the bar strength and dissipation rates balanced. The bar strength is required by self-consistency to be in the weak perturbation regime (less than 5$\%$) but below this threshold there is a degeneracy with the dissipation rates.  The ratio of the two dissipation rates has little effect on the orbits, i.e. there is a degeneracy between them. So these three parameters (bar strength and the two dissipation rates) are not well constrained and only their combination matters.

\section{Resonance radii and pattern speed in a logarithmic potential}\label{app:resonances}

A logarithmic potential is defined as:
\begin{equation}\label{eq:potential}
\Phi_0(r)=\frac{1}{2} v_{\rm p}^2 \ln({1+\frac{r^2}{r_{\rm p}^2}})
\end{equation}
where $v_{\rm p}$ and $r_{\rm p}$ are the characteristic velocity and length of the potential. The rotation frequency is given by:
\begin{equation}\label{eq:rotfreq}
\Omega(r)^2= \frac{1}{r}\frac{\partial \left(\Phi_0\right)}{\partial r}=\frac{v_{\rm p}^2}{r_{\rm p}^{2}+r^{2}}
\end{equation}
and the epicyclic frequency by:
\begin{equation}
\kappa(r)^2=r\* \frac{\partial \left(\Omega^{2}\right)}{\partial r}+4\* \Omega^{2}.
\end{equation}
The  inner Lindblad resonance (ILR) occurs when $\Omega-\kappa/2$ equals the pattern speed of the bar. If we assume only one ILR, this happens when $\Omega-\kappa/2$ is at its maximum.
The radius of the ILR, $R_{\rm ILR}$, is therefore obtained by solving
\begin{equation}
\frac{\partial \left(\Omega-\kappa/2\right)}{\partial r}=0.
\end{equation}
It scales with the scale radius of the potential, $r_{\rm p}$, and is given by
\begin{equation}\label{eq:ri}
r_{\rm i}=R_{\rm ILR}/r_{\rm p}=5^{1/4}
\end{equation}
where $r_{\rm i}$ is the ILR radius in units of $r_{\rm p}$.

The pattern speed of the bar, $\Omega_{\rm p}$, is given by $\Omega_{\rm p}=\Omega(R_{\rm ILR})-\kappa(R_{\rm ILR})/2$:
\begin{equation}\label{eq:pat}
\Omega_{\rm p}=\left(\frac{1}{\sqrt{1+\sqrt{5}}}-\frac{\sqrt{2\* \sqrt{5}+4}}{2\* (\sqrt{5}+1)}\right)\frac{v_{\rm p}}{r_{\rm p}}\simeq 0.11 \frac{v_{\rm p}}{r_{\rm p}}
\end{equation}

The corotation resonance occurs at the radius where the rotational frequency equals the pattern speed of the bar. So, the corotation radius, $R_{\rm c}$, is related to $R_{\rm ILR}$ through 
\begin{equation}
\Omega(R_{\rm c})=\Omega(R_{\rm ILR})-\kappa(R_{\rm ILR})/2,
\end{equation}
which leads to:
\begin{equation}
r_{\rm c}^2=6\* r_{\rm i}^{2}+19+22\* r_{\rm i}^{-2}+8\*r_{\rm i}^{-4}+(4+8\* r_{\rm i}^{-2}+4\*r_{\rm i}^{-4})\* \sqrt{2\* r_{\rm i}^{4}+6\* r_{\rm i}^{2}+4}
\end{equation}
where $r_{\rm c}=R_{\rm c}/r_{\rm p}$ is the corotation radius scaled to  $r_{\rm p}$. 
Replacing $r_{\rm i}$ by Eq.\,\ref{eq:ri} gives:
\begin{equation}\label{eq:rc}
r_{\rm c}=\sqrt{20\* \sqrt{5}+43}.
\end{equation}
The ratio of the resonance radii is given by Eq.\,\ref{eq:ri} and \ref{eq:rc}:
\begin{equation}
R_{\rm c}/R_{\rm ILR}=r_{\rm c}/r_{\rm i}=\frac{\sqrt{20\* \sqrt{5}+43}}{5^{1/4}}\simeq 6.26
\end{equation}

\section{Mass inflow in a logarithmic potential with a bar perturbation}

The non-axisymetric component of the potential (the bar component) is given by
\begin{equation}
\Phi_{\rm b}(r,\phi)= \epsilon\Phi_0 \cos(2\phi),
\end{equation}
where $\Phi_0$ is given by Eq.\,\ref{eq:potential} and $\epsilon$ is the strength of the bar.
The resulting gravitational torque exerted on a particle at $(r,\phi)$ is
\begin{equation}\label{eq:insttorque}
\mathcal{T}(r,\phi)= r F_{\phi}=\frac{\partial \Phi}{\partial \phi}=-\epsilon v_{\rm p}^2 \ln\left({1+\frac{r^2}{r_{\rm p}^2}}\right) \sin(2\phi)
\end{equation}
We define the net torque, $\bar{\mathcal{T}}$, as the average torque over an orbit
\begin{equation}\label{eq:nettorque}
\bar{\mathcal{T}}=  \frac{\int_0^{\tau}\mathcal{T}(t)dt}{\tau},
\end{equation}
where $\tau$ is the orbital period.
The angular momentum variation over this orbit and per unit mass is then given by
\begin{equation}\label{eq:angmomloss}
\Delta \mathcal{M}=\bar{\mathcal{T}}\tau.
\end{equation}
A particle that changes angular momentum will settle on a new orbit which has an intrinsic angular momentum (as determined by the axisymmetrical component of the potential) corresponding to this new value.
Hence, the radial velocity of the matter flow is given by the ratio of this angular momentum variation to the one fixed by the axisymmetric component of the potential and divided by the rotation period.
\begin{equation}
v_{\rm rad} =\frac{\Delta \mathcal{M}}{\partial \mathcal{M}/\partial r} \frac{1}{\tau}=\frac{\bar{\mathcal{T}}}{\partial \mathcal{M}/\partial r}, 
\end{equation}
where $\partial \mathcal{M}/\partial r=\partial (r^2\Omega)/\partial r$. Using Eq.\,\ref{eq:rotfreq} this leads to
\begin{equation}\label{eq:radvel}
v_{\rm rad} =\frac{\bar{\mathcal{T}}}{r\Omega(2-r^2/(r_{\rm p}^2+r^2))}.
\end{equation}

\begin{acknowledgements}
This research has made use of the GOLDMine Database. We thank the scientific and technical staffs at the IRAM 30\,m and PdBI for their help in obtaining the observations presented here. We are grateful to the referee, Eric Emsellem, for helpful comments.
\end{acknowledgements}

\bibliographystyle{aa}
\bibliography{7254}

\end{document}